\documentclass[11pt]{article}
\pdfoutput=1
\usepackage[margin=1.0in]{geometry}
\usepackage{amsmath,amssymb,graphicx}
\usepackage{hyperref}
\usepackage{slashed}

\usepackage[T1]{fontenc}

\usepackage[utf8]{inputenc}

\usepackage[greek,english]{babel}

\usepackage{framed}

\usepackage[font=small,labelfont=bf]{caption}
\usepackage{cite}

\newcommand{\iu}{{\mathrm i}}
\newcommand{\E}{{\mathrm e}}
\newcommand{\cpi}{\text{\greektext p}}

\newcommand{\RR}{{\mathbb{R}}}
\newcommand{\ZZ}{{\mathbb{Z}}}

\newcommand{\rmd}{\textrm{d}}
\newcommand{\dif}[1]{\ensuremath{\operatorname{d}\!{#1}}}

\newcommand{\Uone}{{\mathrm{U(1)}}}
\newcommand{\SU}{{\mathrm{SU}}}

\newcommand{\be}{\begin{equation}}
\newcommand{\ee}{\end{equation}}

\usepackage{xcolor}
\usepackage{bm}

\title{\bf Axion-Gauge Coupling Quantization with a Twist
}
\author{Matthew Reece \\[10pt]
{\small \color{gray} \texttt{mreece@g.harvard.edu}}\\
{\small Department of Physics, Harvard University, Cambridge, MA, 02138, USA}}

\date{September 7, 2023}

\begin{document}

\maketitle

\begin{abstract}

The possible couplings of an axion to gauge fields depend on the global structure of the gauge group. If the Standard Model gauge group is minimal, or equivalently if fractionally charged color-singlet particles are forbidden, then the QCD axion's Chern-Simons couplings to photons and gluons obey correlated quantization conditions. Specifically, the photon coupling can have a fractional part which is a multiple of 1/3, but which is determined by the gluon coupling. A consequence of this result is that, among all theories with a minimal gauge group and minimal axion coupling to gluons, the smallest possible axion-photon amplitude $|g_{a\gamma\gamma}|$ arises for $E/N = 8/3$. This provides a new motivation for experiments targeting this axion-photon coupling.

\end{abstract}

\tableofcontents

\section{Introduction}

The QCD axion is not only the most compelling solution to the Strong CP problem~\cite{Peccei:1977ur, Peccei:1977hh, Wilczek:1977pj, Weinberg:1977ma}, but also an appealing dark matter candidate via the misalignment mechanism~\cite{Preskill:1982cy, Dine:1982ah, Abbott:1982af}. For both of these, only the axion's coupling to gluons plays a direct role. On the other hand, very few of the experimental and observational efforts to detect the axion rely on its gluon coupling (but see~\cite{Graham:2011qk, Budker:2013hfa, Berlin:2022mia, Arvanitaki:2021wjk}); most instead rely on its coupling to photons (e.g.,~\cite{Sikivie:1983ip, ADMX:2006kgb, Graham:2015ouw, Kahn:2016aff, ADMX:2018gho, Chaudhuri:2019ntz, ADMX:2019uok, Berlin:2020vrk, ADMX:2021nhd, DMRadio:2022pkf, Yi:2022fmn}). This makes it crucial to understand how these couplings are related. It is well-known that the coupling of a QCD axion to photons is a sum of two contributions, a quantized piece that depends on UV physics and an additional, non-quantized piece that arises at the QCD scale from mixing with neutral mesons, especially the pion~\cite{Kaplan:1985dv, Srednicki:1985xd, Georgi:1986df, Svrcek:2006yi, Agrawal:2017cmd, GrillidiCortona:2015jxo}. Here we point out an overlooked constraint: the quantized, UV contribution to the axion-photon coupling is nontrivially correlated with the axion-gluon coupling, if the Standard Model gauge group is minimal (or, relatedly, if the UV completion does not involve fractionally charged color-singlet particles). By minimal, we mean the Standard Model gauge group
\begin{equation} \label{eq:GSM}
G_\textsc{SM} \cong [\SU(3)_\textsc{C} \times \SU(2)_\textsc{L} \times \Uone_\textsc{Y}]/\ZZ_6.
\end{equation}
It is possible that this is not the true gauge group: there could be a quotient by $\ZZ_2$ or $\ZZ_3$, or no quotient at all, or there could even exist particles with hypercharge $1/12$ or another fraction smaller than $1/6$. However, this is the minimal possibility, in the sense that it allows the fewest possible representations for electrically charged matter. For this minimal choice of $G_\textsc{SM}$, the axion coupling to photons is not fully independent of the coupling to gluons.

For most of this paper, we focus on just the axion, photon, and gluons. The full Standard Model including the electroweak gauge bosons (and possible non-minimal choices of the gauge group) is discussed in Appendix~\ref{sec:fullSM}; we only briefly comment on axion-fermion couplings in \S\ref{sec:comments}. To fix our conventions: we consider the kinetic terms
\begin{equation}
I_\mathrm{kin} = \int \rmd^4x \sqrt{|g|} \left[\frac{1}{2} f^2 \partial_\mu \theta \partial^\mu \theta - \frac{1}{4 e^2} F_{\mu \nu}F^{\mu \nu} - \frac{1}{2 g_s^2} \mathrm{tr}(G_{\mu \nu}G^{\mu \nu})\right],
\end{equation}
where $\theta \cong \theta + 2\cpi$ is a dimensionless periodic axion field, $F = \rmd A$ is the electromagnetic field strength in the normalization where the electron has charge $-1$, and $G = \rmd C - \iu C \wedge C$ is the gluon field strength (and, for convenience of equations written in terms of differential forms, we make the somewhat uncommon choice to refer to the gluon field as $C_\mu$ so that the gauge field and its field strength are referred to with different letters). We normalize the $\SU(3)$ generators with the standard particle physics convention $\mathrm{tr}(T^a T^b) = \frac{1}{2}\delta^{ab}$. Notice that $\theta$, $A$, and $C$ are {\em not} canonically normalized, but instead are normalized in a manner that makes periodicity and charge quantization properties manifest. Factors of $f$, $e$, and $g_s$ are needed to restore canonical normalization.

The quantities of interest to us are the axion-gauge couplings $k_G$ and $k_F$ in
\begin{equation} \label{eq:Iax}
I_\mathrm{ax} = \int \left[\frac{k_G}{8\cpi^2}\, \theta\, \mathrm{tr}(G \wedge G) + \frac{k_F}{8\cpi^2} \,\theta\, F \wedge F \right]. 
\end{equation}
This action is not invariant under the gauge transformation $\theta \mapsto \theta + 2\cpi$. We require that the path integral be well-defined, so $\exp[\iu I]$ should be gauge invariant. This implies quantization conditions on $k_G$ and $k_F$. In fact, these numbers can be thought of as (generalized) Chern-Simons levels, which are well-known to be quantized. Readers unfamiliar with this physics can find a detailed pedagogical review in my TASI lectures~\cite{Reece:2023czb}. The main result of this paper is that, if the full Standard Model gauge group is the minimal one~\eqref{eq:GSM}, which forbids fractionally charged color-singlet particles, then in fact $k_G$ and $k_F$ are not independently quantized. Instead, we have
\begin{framed}
\begin{equation} \label{eq:mainresult}
k_G \in \ZZ, \qquad \frac{2}{3} k_G + k_F \in \ZZ.
\end{equation}
\end{framed}
The integer $k_G$ is also known as the domain wall number, because QCD dynamics will generate an axion potential with $|k_G|$ distinct vacua that can be separated by domain walls at the QCD phase transition. Eq.~\eqref{eq:mainresult} implies that the photon coupling $k_F$ can be fractional (a multiple of $1/3$), but its fractional part is determined by the domain wall number. This is a nonperturbative fact about the theory, but I will explain how it arises in perturbative theories (e.g., KSVZ-like models) where $k_G$ and $k_F$ are computed from triangle diagrams. In this case, even if the global structure of the Standard Model is non-minimal (e.g., lacking the $\ZZ_6$ quotient), the result still holds provided that the fermions appearing in the UV completion and giving rise to the axion couplings appear in representations of $[\SU(3)_\textsc{C} \times \SU(2)_\textsc{L} \times \Uone_\textsc{Y}]/\ZZ_6$, i.e., do not include fractionally charged color-singlet particles.
 
Experimental searches for axions generally express the result in terms of a quantity $g_{a\gamma\gamma}$, proportional to the axion-photon-photon amplitude at low energies, which is defined to be
\begin{equation}
g_{a\gamma\gamma} = \frac{\alpha}{2\cpi f/k_G} \left(\frac{E}{N} - 1.92(4)\right).
\end{equation}
Here $E \equiv k_F$, $N \equiv \frac{1}{2} k_G$, and the $-1.92$ contribution arises from mixing with mesons (we use a recent estimate from~\cite{GrillidiCortona:2015jxo}). It is very common in the axion literature to refer to a model based on its value of $E/N$, i.e., of $2 k_F/k_G$. The factor of 2 here is a historical artifact, since $k_G$ is an integer but $N$ can be a half-integer. In the literature, what I denote as $f/k_G$ is often denoted $f_a$, and it is this combination that determines the QCD axion's mass, a recent estimate of which is~\cite{GrillidiCortona:2015jxo}:
\begin{equation}
m_a \approx 5.70(6)(4)\,\mu\mathrm{eV} \left(\frac{10^{12}\,\mathrm{GeV}}{f/k_G}\right).
\end{equation}
It is common for experimental plots to show lines in the $(m_a, g_{a\gamma\gamma})$ plane labeled by values of $E/N$, such as $E/N = 0$ (often labeled KSVZ~\cite{Kim:1979if, Shifman:1979if}) or $E/N = 8/3$ (often labeled DFSZ~\cite{Zhitnitsky:1980tq, Dine:1981rt}). We emphasize that the conventional KSVZ and DFSZ labels are misleading, as the structure of these models can accommodate many different $E/N$ values, and conversely the same $E/N$ values can arise in models with completely different structure. The value $E/N = 8/3$ is also of interest because it arises in GUT completions of the Standard Model~\cite{Srednicki:1985xd, Svrcek:2006yi, Agrawal:2022lsp}. Because $E/N = 8/3$ mostly cancels against $-1.92$, and because it arises in simple models, it is often viewed as a major target for axion experiments. Of course, one could consider values like $E/N = 2$ that cancel even more completely, so it is not clear whether $E/N = 8/3$ is a well-motivated stopping point.

Our result provides a different motivation for the line $E/N = 8/3$ as an experimental target. Models with $k_G = \pm 1$ are particularly appealing. A phenomenological corollary of our result~\eqref{eq:mainresult} is:
\begin{framed}
\noindent
If the axion coupling to gluons is minimal ($k_G = 1$), and the Standard Model gauge group is minimal (no fractionally charged color-singlet particles exist), then the smallest value of $|g_{a\gamma\gamma}|$ consistent with eq.~\eqref{eq:mainresult} arises for $k_F = 4/3$, or in conventional notation, $E/N = 8/3$.
\end{framed}
\noindent
Notice that the case $k_G = -1$ also has domain wall number one, but then the smallest $|g_{a\gamma\gamma}|$ arises for $k_F = -4/3$, so the conclusion about $E/N$ and $g_{a\gamma\gamma}$ is unchanged. (In fact, one could always redefine $\theta \mapsto -\theta$ to ensure $k_G > 0$ without affecting $|g_{a\gamma\gamma}|$.)

We give two different arguments for~\eqref{eq:mainresult}. First, in \S\ref{sec:twisted}, we give a nonperturbative argument based on field configurations with nontrivial topology. Second, in \S\ref{sec:reptheory}, we give an argument based on $\SU(3)$ representation theory, which applies to any perturbative model lacking fractionally charged color-singlet particles. We offer further commentary on the result and its implications in \S\ref{sec:comments}. In an appendix \S\ref{sec:fullSM}, we explain how the result changes for different global structures of the Standard Model gauge group.

\section{Argument from twisted field configurations}
\label{sec:twisted}

Given the minimal Standard Model gauge group~\eqref{eq:GSM}, there is a correlation between the $\SU(3)_\textsc{C}$ representations of particles and their electric charges. This correlation depends only on the representation's {\em triality}, or transformation under the $\ZZ_3$ center of $\SU(3)_\textsc{C}$. Representations of zero triality have integer electric charge, those with triality 1 (e.g., the $\bm{3}$) have electric charge $1/3$ less than an integer, and those with triality 2 (e.g., the $\bm{\bar{3}}$) have electric charge $1/3$ more than an integer. 

Below the electroweak scale, we can summarize this result by saying that the gauge group is $G = \mathrm{U}(3) \cong [\SU(3) \times \Uone]/\ZZ_3$. More explicitly, we have charged fields in the fundamental of $\SU(3)$ transforming as 
\begin{equation}
\psi \mapsto U \E^{\iu {\hat q} \alpha} \psi,
\end{equation}
where $U \in \SU(3)$ and $\E^{\iu \alpha} \in \Uone$, with $\hat q$ an integer charge that is 3 times our conventional normalization $q$ of electric charge. For instance, the up quark has $q = 2/3$ and thus $\hat q = 2$. 

The center of $\SU(3)$ is generated by the matrix
\begin{equation} \label{eq:zmatrix}
z = \begin{pmatrix} 
\E^{2\cpi \iu/3} & 0 & 0 \\ 0 & \E^{2\cpi \iu/3} & 0 \\ 0 & 0 & \E^{2\cpi \iu/3}
\end{pmatrix} = \E^{2\cpi \iu/3} \bm{1}.
\end{equation}
and the $\ZZ_3$ quotient corresponds to the fact that the combined action of $z \in \SU(3)$ and the $\Uone$ element with $\alpha = 2\cpi/3$ acts trivially on fields. In order for this to be true for a field in the fundamental, we need 
\begin{equation}
\E^{2\cpi \iu/3 (1 + {\hat q})} = 1,
\end{equation}
i.e., we need
\begin{equation}
{\hat q} \equiv 2 \pmod 3
\end{equation}
for fields in the fundamental of $\SU(3)$. Correspondingly, we need ${\hat q} \equiv 1 \pmod 3$ for fields in the antifundamental of $\SU(3)$.

We will derive our quantization condition using topologically nontrivial field configurations with twisted boundary conditions, of the sort introduced by 't~Hooft~\cite{tHooft:1979rtg, vanBaal:1982ag}. The homotopy group $\pi_1(G) \cong \ZZ$ is generated by the projection of a path in the covering space $\widetilde{G} = \SU(3) \times \Uone$ that goes from the origin $(\bm{1}, 1)$ to the point $(z, \E^{2\cpi \iu/3})$, which is identified with the origin when we take the $\ZZ_3$ quotient. A simple such path $f: [0,1] \to \widetilde{G}$ is given by
\begin{equation} \label{eq:pi1param}
(U(t), \xi(t)) = \left(\exp(2\cpi \iu t T/3), \exp(2\cpi \iu t/3)\right),
\end{equation}
where
\begin{equation} \label{eq:Tmatrix}
T = \begin{pmatrix} 1 & 0 & 0 \\ 0 & 1 & 0 \\ 0 & 0 & -2 \end{pmatrix}
\end{equation}
has the property that $\exp(2\cpi \iu T/3) = z$.

We denote the $\SU(3)$ gauge field by $C$ with field strength $G = \rmd C - \iu C \wedge C$ and the $\Uone$ gauge field by $\widehat{A}$ with field strength $\widehat{F} = \rmd \widehat{A}$, with gauge transformations 
\begin{align}
C &\mapsto U C U^{-1} - \iu (\rmd U) U^{-1} \\
\widehat{A} & \mapsto \widehat{A} - \iu (\rmd \E^{\iu \alpha}) \E^{-\iu \alpha} = \widehat{A} + \rmd \alpha.
\end{align}
We are interested in gauge field configurations that return to themselves with a twist under the $\ZZ_3$ center symmetry. We study the theory on a 4-torus parametrized by $x_i \cong x_i + 2\cpi r_i$. Let's first consider flux on a 2-torus. Consider a gauge field configuration of the form
\begin{equation}
C = G_{12} x_1 \rmd x_2, \quad \widehat{A} = \widehat{F}_{12} x_1 \rmd x_2.
\end{equation}
This is manifestly not invariant under $x_1 \mapsto x_1 + 2\cpi r_1$, but we ask that it map to a gauge equivalent configuration under a gauge transformation that winds around the $x_2$ direction as the generator of $\pi_1(G)$, i.e.,
\begin{align}
2 \cpi r_1 G_{12} \rmd x_2 &= -\iu (\rmd U(x_2/2\cpi r_2)) U^{-1}(x_2/2\cpi r_2), \\
2 \cpi r_1 \widehat{F}_{12} \rmd x_2 &= -\iu \xi^{-1}(x_2/2\cpi r_2) \rmd \xi(x_2/2\cpi r_2).
\end{align}
From the explicit expression~\eqref{eq:pi1param} we read off that
\begin{align}
G_{12} &= \frac{1}{3} \frac{1}{2\cpi r_1 r_2} T, \\
\widehat{F}_{12} &= \frac{1}{3} \frac{1}{2\cpi r_1 r_2}.
\end{align}
Next, consider a field configuration with the same type of flux on both the $(x_1, x_2)$ torus and the $(x_3, x_4)$ torus:
\begin{align}
G &= \frac{1}{3} T \left(\frac{1}{2\cpi r_1 r_2} \dif x_1 \wedge \dif x_2 + \frac{1}{2\cpi r_3 r_4} \dif x_3 \wedge \dif x_4\right), \\
\widehat{F} &= \frac{1}{3} \left(\frac{1}{2\cpi r_1 r_2} \dif x_1 \wedge \dif x_2 + \frac{1}{2\cpi r_3 r_4} \dif x_3 \wedge \dif x_4\right).
\end{align}
Then we calculate that:
\begin{align}
\int \mathrm{tr}(G \wedge G) &= \frac{1}{9} 8 \cpi^2 \mathrm{tr}(T T) = \frac{2}{3} 8 \cpi^2, \\
\int \widehat{F} \wedge \widehat{F} &= \frac{1}{9} 8\cpi^2.
\end{align}
Finally, we note that the conventional normalization of the electromagnetic gauge field is determined by ${\hat q}\widehat{A} = q A = {\hat q} A/3$, hence $A = 3 \widehat{A}$, and
\begin{equation}
\int F \wedge F = 9 \int \widehat{F} \wedge \widehat{F} = 8\cpi^2.
\end{equation}

Now we are ready to put the pieces together. Given the action~\eqref{eq:Iax}, we consider the change in $\exp(\iu I)$ under $\theta \mapsto \theta + 2\cpi$, in the presence of the twisted field configuration we have just considered:
\begin{equation}
\exp(\iu I) \mapsto \exp(\iu I) \exp\left[2\cpi \iu \left(\frac{2}{3} k_G + k_F\right)\right].
\end{equation}
From this we conclude that we require a correlated quantization condition of the form
\begin{equation} \label{eq:U3condition}
\frac{2}{3} k_G + k_F \in \ZZ,
\end{equation}
as promised in~\eqref{eq:mainresult}. The other part of~\eqref{eq:mainresult}, $k_G \in \ZZ$, follows from the standard argument about $\SU(3)$ instantons. 

So far we have shown that~\eqref{eq:mainresult} holds in any valid model with the low-energy gauge group $[\SU(3) \times \Uone]/\ZZ_3$, but one could ask if there might be a stronger condition that holds. In other words, given integers $(n, m)$, is there always a model that produces $k_G = n$ and $\frac{2}{3} k_G + k_F = m$? In a KSVZ-like model, a color triplet $Q$ with electric charge $-\frac{1}{3}$ and PQ charge $\pm 1$ produces $(k_G, k_F) = \pm (1,\frac{1}{3})$, whereas a color singlet $L$ with electric charge $1$ and PQ charge $\pm 1$ produces $(k_G, k_F) = (0, \pm 1)$. A model with sufficiently many copies of such fields can achieve any $(n, m)$; for instance, if $n > 0$, one can choose $n$ copies of $Q$ with PQ charge $+1$ and $|m - n|$ copies of $L$ with PQ charge $\mathrm{sign}(m - n)$. This shows that~\eqref{eq:mainresult} is in fact the most general quantization condition that can be proven.

\section{Argument from representation theory}
\label{sec:reptheory}

In \S\ref{sec:twisted}, we gave a nonperturbative argument for~\eqref{eq:mainresult} exploiting the topology of the gauge group by placing the theory on a topologically nontrivial manifold. We will now show that the same conclusion can be reached within a perturbative model (e.g., a KSVZ-like or DFSZ-like model) by integrating out heavy fermions. This argument has a less general starting point, as it makes assumptions about the UV completion; for example, the axion could arise from a higher-dimensional gauge field, in which case $k_G$ and $k_F$ descend directly from higher-dimensional Chern-Simons levels, which do not arise from integrating out 4d fermions. Nonetheless, it is instructive to see how perturbative models respect the nonperturbative constraint~\eqref{eq:mainresult}, and we will see that in fact we derive a stronger statement: every individual contribution to $k_G$ and $k_F$ independently obeys~\eqref{eq:mainresult}. (This also follows from the topological argument by specializing to a KSVZ model with a single representation of fermions giving rise to the coupling.)

In a perturbative 4d model, the coefficients $k_G$ and $k_F$ arise from triangle diagrams, integrating out particles in $\SU(3)$ representation $R_i$ with electric charge $q_i \in \frac{1}{3} \ZZ$ and Peccei-Quinn charge $p_i \in \ZZ$,
\begin{align}
k_G &= \sum_i 2 I(R_i) p_i, \\
k_F &= \sum_i \dim(R_i) q_i^2 p_i.
\end{align}
Here $I(R_i)$ is the Dynkin index of the representation, normalized to $1/2$ for the fundamental. We will show that each individual contribution to the sum obeys~\eqref{eq:mainresult}, for the minimal choice $|p_i| = 1$ (and thus a fortiori for other choices of $p_i$). Because this holds for every contribution, we will drop the $i$ subscript and focus on a representation $R$; our goal is to argue that
\begin{equation} \label{eq:repclaim}
\frac{4}{3} I(R) + \dim(R) q^2 \in \ZZ,
\end{equation}
for any $\SU(3)$ representation $R$ and $\Uone$ charge $q$ consistent with the gauge group $[\SU(3) \times \Uone]/\ZZ_3$. Rather than trying to give a general argument applicable to other gauge groups, we will give a very direct (but not elegant) analysis of the explicit formulas for the case of interest.

Let us recall some facts about $\SU(3)$ representation theory. An irreducible representation of $\SU(3)$ is labeled by two nonnegative integers (also sometimes referred to as Dynkin indices) $r$, $s$, which correspond to the number of fundamentals and anti-fundamentals that should be tensored together to obtain the representation. These representations have triality $n_3(R) \equiv r - s \pmod 3$. As reviewed in \S\ref{sec:twisted}, the gauge group $[\SU(3) \times \Uone]/\ZZ_3$ has the constraint that the fractional part of the electric charge $q$ is $-n_3(R)/3$. Thus, we can write $q = n - \frac{r - s}{3}$ for some $n \in \ZZ$.

The dimension of an $\SU(3)$ representation is
\begin{equation}
\dim(R) = \frac{1}{2} (r + 1)(s + 1)(r + s + 2).
\end{equation}
If $n_3(R) \neq 0$, then $\dim(R)$ is divisible by 3. For instance, if $n_3(R) = 1$, then $r \equiv s+1 \pmod 3$ and so $2\dim(R) = (s+2)(s+1)(2s+3) \equiv 2s(s+1)(s+2) \pmod 3$, so $\dim(R) \equiv s(s+1)(s+2)\pmod 3$ is a product of 3 consecutive numbers mod 3 and hence is $0 \pmod 3$. By $r \leftrightarrow s$ the same is true if $n_3(R) = 2$. On the other hand, $\dim(R)$ need not be divisible by 3 when $n_3(R) = 0$, as exemplified by the adjoint representation of dimension 8.

The Dynkin index of an $\SU(3)$ representation is
\begin{equation} \label{eq:SU3index}
I(R) = \frac{1}{24} \dim(R) (r^2 + r s + s^2 + 3r + 3 s).
\end{equation}

It is instructive to look at some examples first: 
\begin{itemize}
\item The fundamental representation has $r = 1, s = 0$, $\dim(R) = 3$, and $I(R) = \frac{1}{2}$. Hence $\frac{4}{3} I(R) = \frac{2}{3}$. In this case the allowed electric charges are $q = n - \frac{1}{3}$, so $\dim(R) q^2 = 3 \left(n - \frac{1}{3}\right)^2 = 3 n^2 - 2 n + \frac{1}{3}$. The contribution in~\eqref{eq:repclaim} is then $3 n^2 - 2 n  + 1 \in \ZZ$.

\item The adjoint representation has $r = 1, s = 1$, $\dim(R) = 8$, and $I(R) = 3$. In this case $\frac{4}{3} I(R) = 4$ is an integer, and only integer electric charges are allowed, so~\eqref{eq:repclaim} obviously holds.

\item The symmetric tensor representation has $r = 2, s = 0$, $\dim(R) = 6$, and $I(R) = \frac{5}{2}$. In this case $\frac{2}{3} I(R) = \frac{10}{3}$. The allowed electric charges are $q = n + \frac{1}{3}$, so $\dim(R) q^2 = 6 \left(n + \frac{1}{3}\right)^2 = 6 n^2 + 4 n + \frac{2}{3}$. The two terms in~\eqref{eq:repclaim} sum to $6 n^2 + 4 n + 4 \in \ZZ$.
\end{itemize}

In general, $2I(R) \in \ZZ$. This is physically clear, since we can build a KSVZ model where this is the value of $k_G$, which must be an integer for the usual topological reason related to $\SU(3)$ instantons.

Let's now break down the argument into cases. First, suppose that $n_3(R) = 0$. Then $q \in \ZZ$, so the $k_F$ contribution is an integer, so in order for~\eqref{eq:repclaim} to hold we need $2I(R)$ to be divisible by 3. There is a simple physical argument for this: the representation $R$ could arise in the context of an $\SU(3)/\ZZ_3$ gauge theory, in which case instanton number can have fractional part a multiple of $1/3$. Thus, the coupling $k_F$ in such a theory is quantized in integer multiples of 3. A KSVZ-like model with a single field in representation $R$ must obey this constraint, hence $3 \mid (2I(R))$. One can also check that this follows from~\eqref{eq:SU3index}, along similar lines to the case we discuss below.

Next, consider the case $n_3(R) = 1$. Then the second term in~\eqref{eq:repclaim} has the form
\begin{equation}
\dim(R) \left(n - \frac{1}{3}\right)^2 = \dim(R) n^2 - \frac{2}{3} \dim(R) n + \frac{1}{9} \dim(R).
\end{equation}
Now, we know that $3 \mid \dim(R)$ in this case, so the fractional part comes only from the last term, $\frac{1}{9} \dim(R)$. In other words, when $n_3(R) = 1$, we wish to show that
\begin{equation} \label{eq:subclaim}
\frac{4}{3}I(R) + \frac{1}{9}\dim(R) \in \ZZ.
\end{equation}
After a little algebra, one finds that this is equivalent to claiming that
\begin{equation}
36 \mid (r+1)(s+1)(r+s+2)(r^2 + s^2 + r s + 3 r + 3 s + 2),
\end{equation}
when $r, s$ are nonnegative integers with $r = s+1 \pmod 3$. Thus, we need to find two factors of 3 and two factors of 2 on the right-hand side. We have already argued that $3 \mid (r+1)(s+1)(r+s+2)$. One can also check that, mod 3,
\begin{equation}
r^2 + s^2 + r s + 3 r + 3 s + 2 \equiv (s+1)^2 + s^2 + (s+1)s + 2  \equiv 3s^2 + 3s + 3  \equiv 0 \pmod 3.
\end{equation}
This takes care of our two factors of 3. Next, we want to show that there are two factors of 2. We do this somewhat tediously by checking cases.
\begin{itemize}
\item $r$ odd, $s$ odd: In this case, $(r+1)$, $(s+1)$, and $(r+s+2)$ are all even, so we have three factors of 2 from the $\dim(R)$ factor alone.
\item $r$ even, $s$ even: In this case, $r+s+2$ and also $r^2 + s^2 + r s + 3 r + 3s + 2$ are both even.
\item $r$ odd, $s$ even: In this case, $r+1$ is even, and $r^2 + s^2 + r s + 3r + 3 s + 2 \equiv r^2 + 3 r \equiv 0 \pmod 2$ is even as well.
\item $s$ odd, $r$ even: Just like the last case with $r \leftrightarrow s$.
\end{itemize}

Finally, one can consider the case $n_3(R) = -1$. In this case $q = n + \frac{1}{3}$, but this does not affect the form of~\eqref{eq:subclaim}, and the argument proceeds exactly as above with $r \leftrightarrow s$. 

\section{Comments}
\label{sec:comments}

In this section we collect a series of brief remarks on different aspects of the result~\eqref{eq:mainresult} and its interpretation.

\subsection{Examples}

In the $\SU(5)$ GUT model, the higgsing pattern is $\SU(5) \to [\SU(3)_\textsc{C} \times \SU(2)_\textsc{L} \times \Uone_\textsc{Y}]/\ZZ_6$ through an adjoint vev, so our constraint should apply. The GUT model predicts a relationship~\cite{Srednicki:1985xd, Svrcek:2006yi}
\begin{equation}
k_F = \frac{4}{3} k_G,
\end{equation}
and hence $\frac{2}{3} k_G + k_F = 2 k_G$ is an integer (in fact, an even integer). In this context, the factor of $4/3$ arises as the trace of the square of the matrix
\begin{equation}
\begin{pmatrix} 1/3 & 0 & 0 & 0 & 0 \\ 0 & 1/3 & 0 & 0 & 0 \\ 0 & 0 & 1/3 & 0 & 0 \\ 0 & 0 & 0 & 0 & 0 \\ 0 & 0 & 0 & 0 & -1 \end{pmatrix},
\end{equation}
which embeds the generator of $\Uone$ electromagnetism in $\SU(5)$. This result is most often quoted as $E/N = 8/3$, using the notation explained in the introduction. For more discussion see~\cite{Agrawal:2022lsp}. 

A wide variety of perturbative axion models with matter in different representations of the gauge group $G_\mathrm{SM}$ has been tabulated in~\cite{DiLuzio:2016sbl}, which lists these models in terms of $E/N$ (in our notation, $2k_F/k_G$) and $N_\textsc{DW}$ (in our notation, $k_G$). Thus, our constraint~\eqref{eq:U3condition} can be written in terms of their data as
\begin{equation}
N_\textsc{DW}\left(\frac{2}{3}  + \frac{1}{2} \frac{E}{N}\right) \in \ZZ.
\end{equation}
We have checked that all of their tabulated models obey this constraint. This follows from the general argument in \S\ref{sec:reptheory}, but it is a useful sanity check on our claim.

\subsection{Axion-fermion couplings}

A point that may, at first, bother some readers is that the couplings $k_G$ and $k_F$ are not, by themselves, physical in theories with fermions charged under the gauge group. This is because the amplitude for an axion to interact with gauge fields also depends on the axion's couplings to fermions that can run in loops. The amplitude is a physical invariant, but the individual couplings are not. Indeed, it is often stated that we can freely move the axion coupling between the phase of a fermion mass term, of the form
\begin{equation} \label{eq:fermionmass}
m_\psi \E^{\iu k \theta} \psi \widetilde{\psi} + \mathrm{h.c.},
\end{equation}
and the terms in~\eqref{eq:Iax} using the chiral anomaly. This is true (provided we also keep track of changes in axion derivative couplings under the field redefinition). However, altering the values of $k_G$ and $k_F$ in this way does not change the conclusion~\eqref{eq:mainresult}. In order for~\eqref{eq:fermionmass} to be well-defined, we require that $k \in \ZZ$. Similarly, when we perform an axion-dependent field redefinition to rephase the fermion, e.g., 
\begin{equation}
\psi \mapsto \E^{\iu n \theta} \psi,
\end{equation}
it makes sense only for $n \in \ZZ$. In that case, the chiral anomaly tells us that the field redefinition shifts $k_G$ by $2 I(R) n$ and $k_F$ by $\dim(R) q^2 n$. But this leaves~\eqref{eq:mainresult} untouched, as the argument of \S\ref{sec:reptheory} shows.

\subsection{Axion strings and anomaly inflow}

In \S\ref{sec:reptheory}, we presented a representation theoretic argument as a constraint on perturbative theories with $k_G$ and $k_F$ arising from triangle diagrams with 4d fermions in loops. There is another way to interpret the calculation that applies to other classes of axion theories, such as those where the axion arises from a higher dimensional gauge field.

Any theory of an axion is expected to have axion strings, i.e., dynamical vortices around which $\theta$ winds from 0 to $2\cpi$. (One argument for this is the absence of global symmetries in quantum gravity, because the axion winding number current $\frac{1}{2\cpi} \rmd \theta$ would generate a 2-form global symmetry if axion strings do not exist.) The lack of gauge invariance of Chern-Simons terms has dynamical implications in the presence of various boundaries or defects. For an axion string, in particular, it implies that the string admits chiral zero modes carrying charge under the bulk gauge symmetries. This is the phenomenon of anomaly inflow~\cite{Callan:1984sa}. The charged zero modes in the 2d worldsheet theory must be anomalous under the gauge symmetries, with anomaly coefficients that cancel the bulk inflow terms arising from $k_G$ and $k_F$. The 2d anomaly arises from vacuum polarization diagrams, proportional to $2 I(R)$ for $\SU(3)$ and $\dim(R) q^2$ for $\Uone$. Thus, we can interpret the argument of \S\ref{sec:reptheory} as a constraint on the anomaly coefficients of fermions on the axion string worldsheet in any axion theory, rather than bulk fermions in 4d.

\subsection{Cosmology}

As emphasized by~\cite{DiLuzio:2016sbl, DiLuzio:2017pfr}, axion models with fractionally charged color-singlet particles are severely constrained by experimental results that found the abundance of such particles in our universe must be twenty orders of magnitude below the abundance of ordinary baryons~\cite{Perl:2009zz}. In any model of a post-inflation axion, i.e., one where the universe undergoes a thermal phase transition after inflation that spontaneously breaks a Peccei-Quinn symmetry and produces the QCD axion as a pseudo-Goldstone boson, we expect that all of the particles responsible for generating the axion-gluon and axion-photon couplings are in thermal equilibrium in the early universe. They will then inevitably have a nonzero relic abundance. The tricky part of the relic abundance calculation is understanding whether hadronic effects enhance the annihilation rate after the QCD phase transition enough to suppress the abundance sufficiently; see, e.g.,~\cite{Kang:2006yd,Jacoby:2007nw} for discussions. The result is that annihilation of such particles is not effective enough to reduce their abundance below the experimental bound. Hence, any post-inflation axion model must be free of fractionally charged color-singlet particles, and the constraint~\eqref{eq:mainresult} applies to all such models.

The case $|k_G| = 1$ is of special interest for a post-inflation QCD axion. The post-inflation axion field value is randomized in different parts of the universe. As a result, at the time of the QCD phase transition, domain walls will form interpolating between different minima of the periodic axion potential~\cite{Sikivie:1982qv}. These lead to a cosmological history inconsistent with the universe we observe unless they are somehow dynamically eliminated. The simplest means of eliminating axion domain walls is if axion strings were formed in the early universe, because domain walls can end on cosmic strings~\cite{Kibble:1982ae, Vilenkin:1982ks, Kibble:1982dd}. The number of domain walls ending on a minimal axion string is $|k_G|$. If this is equal to $1$, a single domain wall ends on a string and the entire string-wall network can efficiently annihilate away into radiation~\cite{Everett:1982nm}. If it is larger than one, multiple walls end on a string and the network is frustrated, leading to an inconsistent cosmology. Thus, there is a strong preference in post-inflation axion models for the minimal axion-gluon coupling $|k_G| = 1$.

Summarizing, cosmological considerations for post-inflation axion models point to both the absence of fractionally charged color-singlet particles and a minimal domain wall number $|k_G| = 1$. Given~\eqref{eq:mainresult}, this then implies that the smallest value of $|g_{a\gamma\gamma}|$ is attained for $k_F = 4/3$ and hence $E/N = 8/3$. This provides a model-independent argument for the importance of $E/N = 8/3$ as an experimental target.

\subsection{The experimental target}

A number of experiments are already targeting $E/N = 8/3$ to provide a benchmark small value of $g_{a\gamma\gamma}$. Such experiments are often advertised as having ``DFSZ sensitivity,'' a phrase that I find misleading, though it is widely used and understood. ADMX and CAPP have already achieved this level of sensitivity~\cite{ADMX:2018gho, ADMX:2019uok, ADMX:2021nhd, Yi:2022fmn}. To do so, they assume that axions constitute all of the dark matter in our neighborhood, with an abundance $\rho = 0.45\,\mathrm{GeV}/\mathrm{cm}^3$. However, there is still a substantial, $O(1)$ uncertainty in the local dark matter density~\cite{deSalas:2020hbh, sivertsson2022estimating, Ando:2022kzd}. I would advocate for taking a more conservative approach, using a deliberately low estimate of $\rho$ or targeting a value somewhat below the $|g_{a\gamma\gamma}|$ predicted by $E/N = 8/3$, to be sure that the axion signal is not missed! 

\subsection{Applicability}

The post-inflation QCD axion scenario is just one possibility. It is equally plausible that the axion was already an independent field during inflation. In the compelling class of models where axions arise from extra-dimensional gauge fields (e.g.,~\cite{Witten:1984dg, Choi:2003wr, Svrcek:2006yi, Conlon:2006tq}), there is no Peccei-Quinn phase transition and the axion is intrinsically of the pre-inflation type. In such models, we do not have a clean cosmological argument for the absence of fractionally charged color-singlet particles (they could simply be inflated away, if they were ever produced at all) or for $|k_G| = 1$, as domain walls do not form since inflation leads to a uniform value of the axion field across the universe. Thus, it is somewhat less clear that $E/N = 8/3$ is the appropriate target for a pre-inflation axion.

At the same time, there are still reasons why our underlying assumptions are plausible in a broader class of models. There are many theories in which the spectrum of light matter is a good guide to the global form of the gauge group. This is an active area of investigation in quantum gravity, under the name ``massless charge sufficiency''~\cite{Morrison:2021wuv, Raghuram:2020vxm, Cvetic:2021vsw}. If only very heavy particles carry a particular charge, then the low-energy effective theory has a very good approximate 1-form global symmetry. Similarly, if $|k_G| \neq 1$, then the low-energy effective theory has a good approximate 0-form global symmetry that shifts the axion by multiples of $2\cpi/k_G$. Quantum gravity forbids exact global symmetries and places restrictions on the quality of approximate symmetries, which provides a reason to think that models with the minimal gauge group and minimal value of $|k_G|$ are on a firmer footing, or at least raise fewer questions about the UV completion. Beyond these considerations, minimality is often considered an aesthetically appealing principle.

In short, we cannot claim that any theoretically consistent axion model will have $|g_{a\gamma\gamma}|$ at or below the value it takes for $E/N = 8/3$. However, a class of minimal and hence appealing examples have this property, including post-inflation axion models with the simplest solution to the domain wall problem.

\section*{Acknowledgments}
I thank the theorists of the 1980s for inexplicably leaving basic aspects of axion physics untouched for me to work on in the 2020s. (By writing this I am obviously in danger of receiving messages telling me that I have overlooked a paper that derived exactly~\eqref{eq:mainresult} decades ago; please let me know!) I thank Shu-Heng Shao and Prateek Agrawal for informing me of their related papers appearing simultaneously~\cite{CFLS, AP}. My work is supported in part by the DOE Grant DE-SC0013607.

\appendix

\section{Details for the full Standard Model}
\label{sec:fullSM}

A compact cover of the Standard Model gauge group is
\begin{equation}
\widetilde{G}_\textsc{SM} \cong \SU(3)_\textsc{C} \times \SU(2)_\textsc{L} \times \Uone_\textsc{Y}.
\end{equation}
(The universal cover would have $\RR$ for the last factor, but we don't expect non-compact gauge groups to appear in real-world physics~\cite{Banks:2010zn}.) There are four possible global structures for the gauge group (see, e.g.,~\cite{Tong:2017oea}): $\widetilde{G}_\textsc{SM}$ itself, $\widetilde{G}_\textsc{SM}/\ZZ_2$, $\widetilde{G}_\textsc{SM}/\ZZ_3$, and $G_\textsc{SM} = \widetilde{G}_\textsc{SM}/\ZZ_6$. In the main text we have focused on only the last case, and also have ignored the electroweak structure and discussed only electromagnetism. In this appendix, we will work out the general case. In the cases with a nontrivial quotient, it must be the case that all hypercharges are (in the usual convention) integer multiples of $1/6$. In the case of $\widetilde{G}_\textsc{SM}$, it could be that there are heavy fields with hypercharge a multiple of $1/(6k)$ for $k \neq 1 \in \ZZ$.

As in the main text, we denote the $\SU(3)_\textsc{C}$ gauge fields by $C$ with field strength $G = \rmd C - \iu C \wedge C$. We denote the $\SU(2)_\textsc{L}$ gauge fields as $L$ with field strength $W = \rmd L - \iu L \wedge L$, and the $\Uone_\textsc{Y}$ gauge field by $\widehat{Y}$ with field strength $\widehat{B} = \rmd \widehat{Y}$ in the normalization that the minimal hypercharge is 1, i.e., the hypercharge of the quark doublet field of the Standard Model is 1 (rather than the conventional $1/6$). The conventional hypercharge gauge field strength is then $B = 6 \widehat{B}$. The kinetic terms are
\begin{equation}
I_\mathrm{kin} = \int \rmd^4x \sqrt{|g|} \left[\frac{1}{2} f^2 \partial_\mu \theta \partial^\mu \theta - \frac{1}{2 g_s^2} \mathrm{tr}(G_{\mu \nu}G^{\mu \nu}) -  \frac{1}{2 g^2} \mathrm{tr}(W_{\mu \nu}W^{\mu \nu}) - \frac{1}{4 g'^2} B_{\mu \nu}B^{\mu \nu} \right],
\end{equation}
and the axion couplings of interest are
\begin{equation}
I_\mathrm{ax} = \int \left[\frac{k_G}{8\cpi^2} \,\theta\,\mathrm{tr}(G \wedge G) + \frac{k_W}{8\cpi^2} \,\theta\,\mathrm{tr}(W \wedge W) + \frac{k_B}{8\cpi^2} \,\theta\,B \wedge B\right].
\end{equation}
In our derivation below it will also be useful to normalize the last term as
\begin{equation}
\int \frac{k_{\widehat{B}}}{8\cpi^2} \,\theta\,\widehat{B} \wedge \widehat{B} = \int  \frac{k_B}{8\cpi^2} \,\theta\,B \wedge B \quad \Rightarrow \quad k_{\widehat{B}} = 36 k_B.
\end{equation}

Below the electroweak symmetry breaking scale, we can integrate out the $Z$ boson by setting $L^3 = Y = A$, where $A$ is the conventionally normalized photon field (i.e., normalized so that the electron charge is $-1$, as in the main text). In this case the axion couplings match onto~\eqref{eq:Iax} with the axion-photon coupling
\begin{equation}  \label{eq:kEMfromkEWK}
k_F = \frac{1}{2} k_W + k_B.
\end{equation}

Independent of the global structure of the gauge group, consideration of $\SU(3)_\textsc{C}$ instantons and $\SU(2)_\textsc{L}$ instantons shows that
\begin{equation}
k_G \in \ZZ, \qquad k_W \in \ZZ.
\end{equation}
The nontrivial calculation is to understand the quantization condition involving $k_{\widehat{B}}$ for each of the possible global structures. Here we present the results from the most complex case to the simplest one.

\subsection{Case 1: $G_\textsc{SM} = \widetilde{G}_\textsc{SM}/\ZZ_6$.}

The center of $\SU(2)$ is generated by the matrix $w = -\bm{1}$. The $\ZZ_6$ quotient corresponds to the fact that the combined action of $z \in \SU(3)_\textsc{C}$ (see~\eqref{eq:zmatrix}), $w \in \SU(2)_\textsc{L}$, and the $\Uone_\textsc{Y}$ element with $\alpha = 2\cpi/6$ acts trivially on all of the fields. The fundamental group $\pi_1(G_\textsc{SM}) \cong \ZZ$ is generated by the projection of a path in $\widetilde{G}_\textsc{SM}$ from the origin ($\bm{1}, \bm{1}, 1)$ to the point $(z, w, \E^{2 \cpi \iu/6})$, which we can take to be given by the following map $f: [0,1] \to \widetilde{G}_\textsc{SM}$:
\begin{equation} \label{eq:pi1paramGSM}
(U(t), V(t), \xi(t)) = (\exp(2\cpi \iu t T/3), \exp(\cpi \iu t \sigma), \exp(2\cpi \iu t/6)),
\end{equation}
with $T$ as in~\eqref{eq:Tmatrix} and
\begin{equation}
\sigma = \begin{pmatrix} 1 & 0 \\ 0 & -1 \end{pmatrix}.
\end{equation}

The calculation proceeds along the same lines as in \S\ref{sec:twisted}, only slightly more complicated. Our gauge transformations are
\begin{align}
C &\mapsto U C U^{-1} - \iu (\rmd U) U^{-1}, \\
L & \mapsto V C V^{-1} - \iu (\rmd V) V^{-1}, \\
\widehat{Y} &\mapsto {\widehat Y} + \rmd \xi.
\end{align}
On a 2-torus, we consider a gauge field configuration of the form
\begin{equation}
C = G_{12} x_1 \rmd x_2, \quad L = W_{12} x_1 \rmd x_2, \quad \widehat{Y} = \widehat{B}_{12} x_1 \rmd x_2.
\end{equation}
For this to be well-defined we ask that under $x_1 \mapsto x_1 + 2\cpi r_1$ it map to a configuration that is equivalent under a gauge transformation winding around the $x_2$ direction as the generator of $\pi_1(G_\textsc{SM})$, i.e.,
\begin{align}
2 \cpi r_1 G_{12} \rmd x_2 &= -\iu (\rmd U(x_2/2\cpi r_2)) U^{-1}(x_2/2\cpi r_2), \\
2 \cpi r_1 W_{12} \rmd x_2 &= -\iu (\rmd V(x_2/2\cpi r_2)) V^{-1}(x_2/2\cpi r_2), \\
2 \cpi r_1 \widehat{B}_{12} \rmd x_2 &= -\iu \xi^{-1}(x_2/2\cpi r_2) \rmd \xi(x_2/2\cpi r_2).
\end{align}
Using~\eqref{eq:pi1paramGSM} we read off that
\begin{align}
G_{12} &= \frac{1}{3} \frac{1}{2\cpi r_1 r_2} T, \\
W_{12} &= \frac{1}{2} \frac{1}{2\cpi r_1 r_2} \sigma, \\
\widehat{B}_{12} &= \frac{1}{6} \frac{1}{2\cpi r_1 r_2}.
\end{align}
Next, consider a field configuration with the same type of flux on both the $(x_1, x_2)$ torus and the $(x_3, x_4)$ torus:
\begin{align}
G &= \frac{1}{3} T \left(\frac{1}{2\cpi r_1 r_2} \dif x_1 \wedge \dif x_2 + \frac{1}{2\cpi r_3 r_4} \dif x_3 \wedge \dif x_4\right), \\
W &= \frac{1}{2} \sigma \left(\frac{1}{2\cpi r_1 r_2} \dif x_1 \wedge \dif x_2 + \frac{1}{2\cpi r_3 r_4} \dif x_3 \wedge \dif x_4\right), \\
\widehat{B} &= \frac{1}{6} \left(\frac{1}{2\cpi r_1 r_2} \dif x_1 \wedge \dif x_2 + \frac{1}{2\cpi r_3 r_4} \dif x_3 \wedge \dif x_4\right).
\end{align}
Then we calculate that:
\begin{align}
\int \mathrm{tr}(G \wedge G) &= \frac{1}{9} 8 \cpi^2 \mathrm{tr}(T T) = \frac{2}{3} 8 \cpi^2, \\
\int \mathrm{tr}(W \wedge W) &= \frac{1}{4} 8 \cpi^2 \mathrm{tr}(\sigma \sigma) = \frac{1}{2} 8 \cpi^2, \\
\int \widehat{B} \wedge \widehat{B} &= \frac{1}{36} 8\cpi^2.
\end{align}
Invariance of $\exp(i I)$ then translates into the correlated quantization condition
\begin{equation}
\widetilde{G}_\textsc{SM}/\ZZ_6: \qquad \frac{2}{3} k_G + \frac{1}{2} k_W + \frac{1}{36} k_{\widehat{B}} = \frac{2}{3} k_G + \frac{1}{2} k_W + k_B  \in \ZZ.
\end{equation}
Now, because of~\eqref{eq:kEMfromkEWK}, this is identical to the condition~\eqref{eq:U3condition} that we derived in the main text from the $\mathrm{U}(3)$ symmetry below the electroweak scale.

\subsection{Case 2: $\widetilde{G}_\textsc{SM}/\ZZ_3$.}

In this case, the fundamental group is generated by the projection of a path in $\widetilde{G}_\textsc{SM}$ from $(\bm{1}, \bm{1}, 1)$ to $(z, \bm{1}, \E^{2\cpi \iu/3})$. We can take this path to be stationary in the $\SU(2)_\textsc{L}$ factor, and then the calculation is precisely like what we discussed in \S\ref{sec:twisted} except that the $\Uone$ factor we are focusing on now is hypercharge rather than electromagnetism. As a result, one derives
\begin{equation}
\widetilde{G}_\textsc{SM}/\ZZ_3: \qquad \frac{2}{3} k_G + \frac{1}{9} k_{\widehat{B}} = \frac{2}{3} k_G + 4 k_B \in \ZZ.
\end{equation}
For example, one could consider a configuration with $k_G = 1, k_W = 0$, and then an allowed choices is $k_{\widehat{B}} = -6$. Below the electroweak scale, this leads to an axion-photon coupling of $k_F = -\frac{1}{6}$. Such a value, which is not an integer multiple of $1/3$, can never arise for the minimal Standard Model gauge group.

\subsection{Case 3: $\widetilde{G}_\textsc{SM}/\ZZ_2$.}

In this case, the fundamental group is generated by the projection of a path in $\widetilde{G}_\textsc{SM}$ from $(\bm{1}, \bm{1}, 1)$ to $(\bm{1}, w, \E^{\cpi \iu})$. Taking the path to be stationary in $\SU(3)_\textsc{C}$ and following the familiar logic, we consider the field configurations
\begin{align}
W &= \frac{1}{2} \sigma \left(\frac{1}{2\cpi r_1 r_2} \dif x_1 \wedge \dif x_2 + \frac{1}{2\cpi r_3 r_4} \dif x_3 \wedge \dif x_4\right), \\
\widehat{B} &= \frac{1}{2} \left(\frac{1}{2\cpi r_1 r_2} \dif x_1 \wedge \dif x_2 + \frac{1}{2\cpi r_3 r_4} \dif x_3 \wedge \dif x_4\right).
\end{align}
Then we find that 
\begin{align}
\int \mathrm{tr}(W \wedge W) &= \frac{1}{4} 8 \cpi^2 \mathrm{tr}(\sigma \sigma) = \frac{1}{2} 8 \cpi^2, \\
\int \widehat{B} \wedge \widehat{B} &= \frac{1}{4} 8\cpi^2,
\end{align}
and finally a correlated quantization condition
\begin{equation}
\widetilde{G}_\textsc{SM}/\ZZ_2: \qquad \frac{1}{2} k_W + \frac{1}{4} k_{\widehat{B}} = \frac{1}{2} k_W + 9 k_B \in \ZZ.
\end{equation}
For example, we could take $k_W = 1$, $k_{\widehat{B}} = -2$, and then $k_F = \frac{1}{2} - \frac{1}{18} = \frac{4}{9}$. Again, we see that this larger gauge group allows for a wider range of axion-photon couplings than in the minimal theory.

\subsection{Case 4: $\widetilde{G}_\textsc{SM}$.}

In the case where there is no quotient, we have only the usual configurations of pure $\Uone_\textsc{Y}$ flux to think about. In this case, because we have normalized $\widehat{B}$ so that the minimum hypercharge is 1, it is a standard result that
\begin{equation}
k_{\widehat{B}} = 36 k_B \in \ZZ.
\end{equation}
This allows the axion-photon coupling $k_F$ to be as small as $\frac{1}{36}$, much smaller than it can be for the minimal Standard Model gauge group.

More generally, we can consider the case that the gauge group is $\widetilde{G}_\textsc{SM}$ but with an even smaller minimal hypercharge $q_\mathrm{min}$ (in the usual normalization of hypercharge). In this case
\begin{equation}
\widetilde{G}_\textsc{SM}: \qquad \frac{1}{q_\mathrm{min}^2} k_B \in \ZZ,
\end{equation}
and the axion-photon coupling can be a multiple of $q_\mathrm{min}^2$.

\bibliography{ref}
\bibliographystyle{utphys}

\end{document}